\documentclass[12pt]{iopart}

\begin{document}

\title[]{Quantum Gravity Effect on the Tunneling Particles from 2+1 dimensional New-type Black Hole}

\author{Ganim Gecim and Yusuf Sucu}

\address{Department of Physics, Faculty of Science, Akdeniz University, 07058 Antalya, Turkey}
\ead{gecimganim@gmail.com and ysucu@akdeniz.edu.tr}

\vspace{10pt}
%\begin{indented}
%\item[]February 2014
%\end{indented}

\begin{abstract}
We investigate the Generalized Uncertainty Principle (GUP) effect on the
Hawking temperature for the 2+1 dimensional New-type black hole by using the
quantum tunneling method for both the spin-1/2 Dirac and the spin-0 scalar
particles. In computation of the GUP correction for the Hawking temperature of the
black hole, we modified Dirac and Klein-Gordon equations. We observed that the
modified Hawking temperature of the black hole depends not only on the black
hole properties, but also on the graviton mass and the intrinsic properties of the
tunneling particle, such as total angular momentum, energy and mass. Also,
we see that the Hawking temperature was found to be probed by these particles in different manners. The modified Hawking temperature for the
scalar particle seems to be lower compared to its standard Hawking temperature. Also, we find that the modified Hawking temperature of the black hole caused by Dirac particle's tunnelling rised by the total angular momentum of the particle. It is diminishable by the energy and mass of the particle and graviton mass as well. These intrinsic properties of the particle, except total angular momentum for the Dirac particle, and graviton mass may cause screening for the black hole radiation.
\end{abstract}

% Uncomment for PACS numbers
%\pacs{00.00, 20.00, 42.10}
%
% Uncomment for keywords
%\vspace{2pc}
%\noindent{\it Keywords}: XXXXXX, YYYYYYYY, ZZZZZZZZZ
%
% Uncomment for Submitted to journal title message
%\submitto{\JPA}
%
% Uncomment if a separate title page is required
%\maketitle
%
% For two-column output uncomment the next line and choose [10pt] rather than [12pt] in the \documentclass declaration
%\ioptwocol
%

\section{Introduction} \label{intro}

Black hole radiation is theoretically very important phenomenon for researchers who attempts to merge the gravitation with the thermodynamics and the quantum mechanics \cite{greif,carter,beken1,beken2,bar,1,1a,1b}. With the
formulation of the quantum field theory in curved spacetime based on
the framework of the standard Heisenberg uncertainty principle, it was
proved that a black hole can emit particles that are created by the quantum
vacuum fluctuation near its outer horizon \cite{1,1a,1b}. Since then, many
alternative methods have been proposed to derive the black hole radiation
known as Hawking radiation in the literature. For instance, the
semi-classical method, based on quantum tunneling process of a particle
across the outer horizon of a black hole from inside to outside, can be used to derive the Hawking radiation. The method implies two different approaches
to compute the imaginary part of the action ($S$), which is the classically
forbidden trajectory of a particle across the outher horizon: the null
geodesic \cite{kraus1,kraus2,par1,par2} and the Hamilton-Jacobi \cite{H1,H2,H3}. In both approaches, the tunneling probability of a
particle from a black hole, $\Gamma $, is defined in terms of the classical
action, as $\Gamma =e^{-\frac{2}{\hbar }ImS}$ \cite{kraus1,kraus2,par1,par2,H1,H2,H3}. By using the semi-classical method, a
lot of studies about the Hawking radiation of a black hole as quantum
tunneling process of a point-like particle have been carried out in the
literature \cite{sa,yale,kerner1,kerner2,kerner3,GY0,GY1,GY2,18,18aa,18a,18b,18c,19,19a,19c,19d,19e,GY3}.
On the other hand, above mentioned studies on this method provide no specific information on type of the particle that is tunnelled from a black hole. That is because the Hawking radiation does not dependent on the intrinsic properties, such as mass, total (orbital+spin) angular momentum, energy and charge, of the tunneling point-like particle.

The existence of a minimal observable length which can be identified by
the order of the Planck scale is a characteristic of the candidate theories
of quantum gravity, such as string theory, loop quantum gravity and
noncommutative geometry \cite{kem1,kem2,ali,das,carr}. This length lead us to a generalized uncertainty principle (GUP) instead of the
standard Heisenberg uncertainty principle. Because a particle is not a
point-like particle in the context of these candidate theories anymore. Therefore, the uncertainty on the
momentum of a particle increases and thus the standard Heisenberg
uncertainty principle can be generalized as follows;
\begin{eqnarray}
\Delta x\Delta p\geq \frac{\hbar}{2}\left[1+\beta (\Delta
p)^{2}\right],\label{GUP1}
\end{eqnarray}
where $\beta=\beta_{0}/ M_{p}^2$, the $M_{p}^2$ is the Planck mass, $\beta_{0}$ is the dimensionless parameter \cite{24a,24b,hos,faizal}. The commutation relations between a particle position, $x$, and momentum, $p$, are modified in the following way;
\begin{eqnarray}
\left[x_{i},p_{j}\right]=i\hbar\delta_{ij}\left[1+\beta p^{2}\right],\label{GUP2}
\end{eqnarray}
where $x_{i}$ and $p_{j}$ represent the modified position and momentum
operators, respectively, and their definitions are as follows;
\begin{eqnarray}
x_{i} &=&x_{0i}, \nonumber \\
p_{j} &=&p_{0j}(1+\beta p_{0}^{2}). \label{GUP3}
\end{eqnarray}
The $x_{0i}$ and $p_{0j}$=$-i\hbar \partial _{j}$ in Eq.(\ref{GUP3}) are the standard
position and momentum operators, respectively, and $p_{0}^{2}$=$p_{0j}p^{0j}$ \cite{24f}. Then, the modified energy expression becomes
\begin{eqnarray}
\widetilde{E}=E\left( 1-\beta E^{2}\right) =E\left[ 1-\beta \left(p^{2}+m_{0}^{2}\right) \right], \label{EN}
\end{eqnarray}
for which the energy mass shell condition, $E^{2}$=$p^{2}+m_{0}^{2}$, is used. From
these relations, the square of the momentum operator can be derived by the
following way,
\begin{eqnarray}
p^{2}=p_{i}p^{i}\simeq-\hbar^{2}\left[ \partial _{i}\partial
^{i}-2\beta \partial _{j}\partial^{j} \left( \partial _{i}\partial ^{i}\right)\right],\label{MO}
\end{eqnarray}
where the higher order terms of the $\beta$ parameter are neglected.

The GUP relations are of great help to understand the nature of a black hole since quantum effects are the essential effects near the event horizon of a black hole. Recently, to
investigate the quantum effects under the GUP relations, the thermodynamics
properties of various black holes have been studied by using the quantum
tunneling process of particles with various spins \cite{noz,jan,24aa,GWS,24c,24d,24e,24f,24g,24h,24j,24k,24l,24m,24n,liz,feng,GY4,GY5}. These studies indicate that the modify Hawking radiation depends not only on the black hole's properties but also on the intrinsic properties of the
tunneling particle.

The New-type black hole is one of the important results of the New Massive Gravity, which is $2+1$ dimensional gravity and graviton in this theory has a mass \cite{kwon}. In the framework of the standard Heisenberg uncertainty
principle, the Hawking radiation of the New-type black holes had been studied
by using the quantum tunneling process of the scalar, Dirac and vector boson
particles \cite{GY1,GY3}. In these studies, it was shown that the Hawking
radiation only depends on the properties of the black hole and is independent
from the properties of the tunneling point-particles, that is, all these
particles tunnel from the black holes in the same way. This indicate that, even if an observer in enough or safe (i.e infinite) distant from
a black hole may detect Hawking radiation of the black hole, the observer can not
determine what kind of particles compose of the radiation. Therefore, in this
study, we will investigate whether the properties of the tunneling particles
will affect the Hawking radiation of the black hole by using the quantum
tunneling process of both the scalar and Dirac particles in the framework of
the GUP.

The organization of this work are as follows: In the Section 2, we modify
the Dirac equation with respect to the GUP relations. Subsequently, using
the modified Dirac equation, we calculate the tunneling possibility of the
Dirac particle by using the Hamilton-Jacobi method, and, then, we find the
modify Hawking temperature of the black hole. In the Section 3, the standard
Klein-Gordon equation is rewritten under the GUP for the $2+1$ dimensional
New-type black hole. Subsequently, the tunneling probability of the scalar
particle from the black hole and the modify Hawking temperature of the black
hole are calculated, respectively. In conclusion, we evaluate and summarize
the results.

\section{Dirac particle's tunneling in the New-type Black Hole}\label{Dirac}

Using the GUP relations, the standard Dirac equation given in Ref. \cite{sucu} can be modified as follows;
\begin{eqnarray}
-i\overline{\sigma }^{0}(x)\partial _{0}\widetilde{\Psi} =\left(
i\overline{\sigma }^{i}(x)\partial _{i}-i\overline{\sigma
}^{\mu}(x)\Gamma_{\mu }-\frac{m_{0}}{\hbar}\right) \left( 1+\beta
\hbar^{2}\partial _{j}\partial ^{j}-\beta m_{0}^{2}\right)
\widetilde{\Psi}, \label{MDE1}
\end{eqnarray}
and its explicit form is
\begin{eqnarray}
i\overline{\sigma }^{0}(x)\partial_{0}\widetilde{\Psi}+i\overline{\sigma
}^{i}(x)\left( 1-\beta m_{0}^{2}\right)\partial_{i}\widetilde{\Psi}+i\beta
\hbar^{2}\overline{\sigma }^{i}(x)\partial _{i}\left(\partial
_{j}\partial^{j}\widetilde{\Psi}\right)-\frac{m_{0}}{\hbar}\left(1-\beta m_{0}^{2}\right)\widetilde{\Psi} \nonumber \\-m_{0}\beta
\hbar\partial_{j}\partial ^{j}\widetilde{\Psi}
-i\overline{\sigma }^{\mu}(x)\Gamma_{\mu}\left( 1+\beta
\hbar^{2}\partial _{j}\partial ^{j}-\beta
m_{0}^{2}\right)\widetilde{\Psi} =0,\label{MDE2}
\end{eqnarray}
where the $\widetilde{\Psi}$ is the modify Dirac spinor, $m_{0}$ is mass of the Dirac particle, $\overline{\sigma }^{\mu }(x)$ are the spacetime dependent Dirac matrices, the $\Gamma_{\mu }(x)$ are the spin affine connection for spin-1/2 particle \cite{sucu}. The spacetime background of
the New-type black hole is given by
\begin{equation}
ds^{2}=L^{2}\left[ f\left( r\right) dt^{2}-\frac{1}{f\left( r\right) }%
dr^{2}-r^{2}d\phi ^{2}\right] ,  \label{Equation1}
\end{equation}%
where $L$ is the AdS$_{3}$ radius defined as $L^{2}$=$\frac{1}{2m^{2}}$=$%
\frac{1}{2\Lambda }$ where $\Lambda $ is cosmological constant and $m$ is
graviton mass, and $f(r)$=$(r-r_{+})(r-r_{-})$ is defined in terms of the
outer, $r_{+}$, and inner, $r_{-}$, horizons radius, respectively. The black
hole's horizons are located at $r_{\pm }=\frac{1}{2}(-b\pm \sqrt{b^{2}-4c})$, where $b$ and $c$ are two constant parameters \cite{GY1,kwon}. Using the Eq.(\ref{Equation1}), the spinorial affine connections are derived
as follows \cite{GY1};
\begin{eqnarray}
\Gamma _{0}=-\frac{i}{4}f^{^{\prime }}(r){\sigma}^{3}{\sigma }^{1}\ ,\ \ \Gamma _{1}=0\ ,\ \Gamma_{2}=\frac{1}{2}\sqrt{f(r)}{\sigma }^{1}{\sigma}^{2}. \label{connec}
\end{eqnarray}

To calculate the tunneling probability of a Dirac particle from the black
hole, we use the following ansatz for the modified wave function;
\begin{equation}
\widetilde{\Psi}(x)=\exp \left( \frac{i}{\hbar }S\left( t,r,\phi
\right) \right)\ \left(\begin{array}{c}A\left( t,r,\phi \right)
\\B\left( t,r,\phi \right) \\ \end{array}\right) \label{Equation13}
\end{equation}
where the $A\left( t,r,\phi \right) $ and $B\left( t,r,\phi \right) $ are
the functions of space-time. The $S(t,r,\phi )$ is the classical action term
for particle trajectory. Inserting the Eqs.(\ref{connec}) and (\ref{Equation13}) in Eq.(\ref{MDE2}), we obtain the following equations for the leading order in $\hbar $ and $\beta $ as
\begin{eqnarray*}
A\left[ \frac{1}{L\sqrt{f}}\frac{\partial S}{\partial t}+m_{0}\left(
1-\beta m_{0}^{2}\right) +\frac{\beta m_{0}}{L^{2}r^{2}}\left( \frac{%
\partial S}{\partial \phi}\right) ^{2}+\frac{\beta m_{0}f}{L^{2}}\left( \frac{%
\partial S}{\partial r}\right) ^{2}\right]  \\
+B\left[ i\frac{\sqrt{f}\left( 1-\beta m_{0}^{2}\right)
}{L}\frac{\partial
S}{\partial r}+\frac{\left( 1-\beta m_{0}^{2}\right) }{Lr}\frac{\partial S}{%
\partial \phi}+i\frac{\beta f^{3/2}}{L^{3}}\left( \frac{\partial S}{\partial r}%
\right) ^{3}\right]\\
+B\left[ i\frac{\beta \sqrt{f}}{L^{3}r^{2}}\frac{\partial S}{\partial r}%
\left( \frac{\partial S}{\partial \phi}\right) ^{2}+\frac{\beta f}{L^{3}r}\frac{%
\partial S}{\partial \phi}\left( \frac{\partial S}{\partial r}\right) ^{2}+%
\frac{\beta }{L^{3}r^{3}}\left( \frac{\partial S}{\partial \phi}\right) ^{3}%
\right]=0,
\end{eqnarray*}
\begin{eqnarray}
A\left[ -i\frac{\sqrt{f}\left( 1-\beta m_{0}^{2}\right)
}{L}\frac{\partial
S}{\partial r}+\frac{\left( 1-\beta m_{0}^{2}\right)}{L r}\frac{\partial S}{%
\partial \phi}-i\frac{\beta f^{3/2}}{L^{3}}\left( \frac{\partial S}{\partial r}%
\right) ^{3}\right] \nonumber  \\
+A\left[ -i\frac{\beta \sqrt{f}}{L^{3}r^{2}}\frac{\partial S}{\partial r}%
\left( \frac{\partial S}{\partial \phi}\right) ^{2}+\frac{\beta f}{L^{3}r}\frac{%
\partial S}{\partial \phi}\left( \frac{\partial S}{\partial r}\right) ^{2}+%
\frac{\beta }{L^{3}r^{3}}\left( \frac{\partial S}{\partial \phi}\right) ^{3}%
\right] \nonumber \\
+B\left[ \frac{1}{L\sqrt{f}}\frac{\partial S}{\partial
t}-m_{0}\left(
1-\beta m_{0}^{2}\right) -\frac{\beta m_{0}}{L^{2}r^{2}}\left( \frac{%
\partial S}{\partial \phi}\right) ^{2}-\frac{\beta m_{0}f}{L^{2}}\left( \frac{%
\partial S}{\partial r}\right) ^{2}\right]=0. \label{MEqq7}
\end{eqnarray}
These two equations have nontrivial solutions for the $A\left( t,r,\phi
\right) $ and $B\left( t,r,\phi \right) $ when the determinant of the
coefficient matrix is vanished. Accordingly, neglecting of the terms
containing higher order of the $\beta$ parameter, provides
\begin{eqnarray}
\beta \left[ 2m_{0}^{4}-\frac{4f}{L^{4}r^{2}}\left( \frac{\partial S}{%
\partial r}\right) ^{2}\left( \frac{\partial S}{\partial \phi }\right) ^{2}-%
\frac{2}{L^{4}r^{4}}\left( \frac{\partial S}{\partial \phi }\right) ^{4}-%
\frac{2f^{2}}{L^{4}}\left( \frac{\partial S}{\partial r}\right)
^{4}\right]\nonumber  \\+\frac{1}{L^{2}f}\left( \frac{\partial S}{\partial t}\right) ^{2}-\frac{f}{%
L^{2}}\left( \frac{\partial S}{\partial r}\right) ^{2}-\frac{1}{L^{2}r}%
\left( \frac{\partial S}{\partial \phi }\right) ^{2}-m_{0}^{2}=0.
\label{MEqq8}
\end{eqnarray}
Due to the commuting Killing vectors $\left(\partial_{t}\right)$ and $\left(\partial_{\phi}\right)$, we can separate the $S\left(t,r,\phi\right)$, in terms of the variables $t$, $r$ and $\phi$, as $S\left(
t,r,\phi \right) =-Et+j\phi +K(r)$, where, $E$ and $j$ are the energy and
angular momentum of the particle, respectively, and $K(r)=K_{0}(r)+\beta
K_{1}(r)$ \cite{24k}. Using these definition in Eq.(\ref{MEqq8}), the
integral of the radial equation, $K(r)$, becomes
\begin{equation}
K_{\pm }(r)=\pm \int \frac{\sqrt{E^{2}-f(r)\left(
m_{0}^{2}L^{2}+j^{2}/r^{2}\right) }}{f(r)}\left[ 1+\beta \chi\right]
dr, \label{MEqq9}
\end{equation}
where $\chi$ is an abbreviation and it is
\begin{equation*}
\chi=\frac{E^{2}\left[ E^{2}-2m_{0}^{2}L^{2}f(r)\right]
}{L^{2}f(r)\left[ E^{2}-f(r)\left( m_{0}^{2}L^{2}+j^{2}/r^{2}\right)
\right]}.
\end{equation*}
Then, by integrating the radial equation, $K_{\pm }(r)$ are obtained as
\begin{equation}
K_{\pm }(r)=\pm i\pi \frac{E}{r_{+}-r_{-}}\left[ 1+\beta\Pi\right],\label{MEqq10}
\end{equation}
where the abbreviation $\Pi$ is
\begin{equation*}
\Pi=\frac{\left(r_{+}-r_{-}\right)^{2}\left[3L^{2}m_{0}^{2}r_{+}^{2}-j^{2}\right] +4E^{2}r_{+}^{2}}{2L^{2}r_{+}^{2}\left( r_{+}-r_{-}\right)^{4}}.
\end{equation*}
On the other hand, the tunneling probabilities of particles
crossing the outer horizon are given by
\begin{eqnarray}
P_{out}=\exp \left[ -\frac{2}{\hbar} ImK_{+}\left( r\right)\right],
\nonumber \\
P_{in}=\exp \left[ -\frac{2}{\hbar}ImK_{-}\left( r\right)
\right]. \label{Equation16}
\end{eqnarray}
Hence, the tunneling probability of the Dirac particle is given by
\begin{equation}
\Gamma=e^{-\frac{2}{\hbar }ImS}=\frac{P_{out}}{P_{in}}=\exp \left( -\frac{4\pi E}{\hbar\left(
r_{+}-r_{-}\right) }\left[ 1+\beta \Pi\right]
\right),\label{MEqq11}
\end{equation}
where $ImS\left(t,r,\phi \right)$=$ImK_{+}\left( r\right) -ImK_{-}\left( r\right)$ \cite{17,17a} and $ImK_{+}\left( r\right)$=$-ImK_{-}\left( r\right)$.
Then, the modify Hawking temperature of the Dirac particle, $T_{H}^{D}$, is obtained as
\begin{equation}
T_{H}^{D}=\frac{\hbar\left( r_{+}-r_{-}\right) }{4\pi}\frac{1}{\left[ 1+\beta \Pi_{D}\right]},\label{MEqq12}
\end{equation}
where to find the temperature it is used the the following relation \cite{cri,vol1,vol2,vol3}:
\begin{equation}
\Gamma =e^{-\frac{2}{\hbar}ImS}=e^{-\frac{E}{T_{H}}}.
\label{Equation18}
\end{equation}
If the $T_{H}^{D}$ are at first expanded in terms of the $\beta$ powers and second
neglected the higher order of the $\beta$ terms, then the modify Hawking
temperature of the black hole is obtained as
\begin{equation}
T_{H}^{D}\simeq\frac{\hbar\left( r_{+}-r_{-}\right) }{4\pi}\left[ 1-\beta \Pi\right] \label{MEqq13}
\end{equation}
From these results, we see that the modified Hawking temperature includes not
only the mass parameter of the black hole, but also the AdS$_{3}$ radius, $L$, (and, hence, the graviton mass) and the angular momentum, energy and mass of the tunnelled Dirac
particle. On the other hand, in the case of $\beta =0$, the modified Hawking temperature is reduced to the
standard temperature obtained by quantum tunneling process of the point particles
with spin-0, spin-1/2 and spin-1, respectively \cite{GY1,GY3}.

\section{Scalar particle's tunneling in the New-type Black Hole}\label{Scalar}

To investigate the quantum gravity effects on the tunneling process of the
scalar particles from the black hole, by using the GUP relations, the
standard Klein-Gordon equation are modified as
\begin{eqnarray}
-\left( i\hbar\right) ^{2}\partial _{t}\partial ^{t}\widetilde{\Phi} =\left[
\left( -i\hbar\right) ^{2}\partial _{i}\partial
^{i}-M_{0}^{2}\right] \left[ 1-2\beta \left( -\hbar^{2}\partial
_{i}\partial ^{i}+M_{0}^{2}\right) \right] \widetilde{\Phi}, \label{MKG1}
\end{eqnarray}
and its explicit form of the modified Klein-Gordon equation is written as
follows;
\begin{eqnarray}
\hbar^{2}\partial_{t}\partial^{t}\widetilde{\Phi}+ \hbar^{2}\partial_{i}\partial^{i}\widetilde{\Phi}+ 2\beta \hbar^{4}\partial_{i}\partial^{i}(\partial_{i}\partial^{i}\widetilde{\Phi}) +M_{0}^2 (1-2\beta M_{0}^2)\widetilde{\Phi}=0, \label{MKG01}
\end{eqnarray}
where $\widetilde{\Phi}$ and $M_{0}$ are the modify wave function and mass of the scalar
particle, respectively. Then, the modified Klein-Gordon equation in the
New-type black hole background becomes as follows:
\begin{eqnarray}
\frac{\hbar^{2}}{f}\frac{\partial ^{2}\widetilde{\Phi }}{\partial t^{2}}-\frac{\hbar^{2}}{r^{2}}%
\frac{\partial ^{2}\widetilde{\Phi }}{\partial \phi^{2}}-2\beta \hbar^{4}f\frac{%
\partial ^{2}}{\partial r^{2}}\left[ -\frac{f}{L^{2}}\frac{\partial ^{2}%
\widetilde{\Phi }}{\partial r^{2}}\right] -\frac{2\beta \hbar^{4}}{r^{2}}\frac{%
\partial ^{2}}{\partial \phi^{2}}\left[ -\frac{1}{L^{2}r}\frac{\partial ^{2}%
\widetilde{\Phi }}{\partial \phi^{2}}\right]\ \nonumber \\ -\hbar^{2}f%
\frac{\partial ^{2}\widetilde{\Phi }}{\partial r^{2}}
+M_{0}^{2}L^{2}\left( 1-2\beta M_{0}^{2}\right) \widetilde{\Phi }=0.
\label{ScalarD2}
\end{eqnarray}
To consider the tunneling radiation of the black hole with the Eq.(\ref{ScalarD2}), we employ the modify wave function of the scalar particle as,
\begin{eqnarray}
\widetilde{\Phi}\left(t,r,\phi \right)=A\exp \left(\frac{i}{\hbar }S\left( t,r,\phi \right) \right), \label{ansatz}
\end{eqnarray}
where $A$ is a constant. Substituting the Eq.(\ref{ansatz}) into the Eq.(\ref{ScalarD2}) and neglecting the the higher order terms of $\hbar $, we get
the equation of motion of the scalar particle as
\begin{eqnarray}
\left(\frac{\partial S}{\partial t}\right)^{2}-f(r)^{2}\left(
\frac{\partial S}{\partial
r}\right)^{2}-\frac{f(r)}{r^{2}}\left(\frac{\partial S}{\partial
\phi}\right)^{2}-M_{0}^{2}L^{2}f(r)-\beta \frac{2f(r)}{r^{4}L^{2}}
\left(\frac{\partial S}{\partial \phi}\right)^{4} \nonumber
\\ +\beta \left[M_{0}^{4}L^{2}f(r)-\frac{2f(r)^{3}}{L^{2}%
}\left(\frac{\partial S}{\partial r}\right) ^{4}\right]=0
\label{HamiltonM}
\end{eqnarray}
Using $S\left( t,r,\phi \right) =-Et+j\phi +W(r)$, where $E$ and $j$ are the
energy and angular momentum of the particle, respectively, and $W(r)=W_{0}(r)+\beta W_{1}(r)$ \cite{24k}, then, the radial integral, $W(r)$,
becomes as follows;
\begin{equation}
W_{\pm }(r)=\pm \int \frac{\sqrt{E^{2}-f(r)\left(M_{0}^{2}L^{2}+j^{2}/r^{2}\right) }}{f(r)}\left[ 1+\beta \Omega\right]dr,
\label{MEqq1}
\end{equation}
where the abbreviation $\Omega$ is
\begin{equation*}
\Omega=\frac{f(r)^{2}\left(M_{0}^{4}L^{4}-j^{4}/r^{4}\right)
-\left[E^{2}-f(r)\left( M_{0}^{2}L^{2}+j^{2}/r^{2}\right) \right]
^{2}}{L^{2}f(r)\left[E^{2}-f(r)\left(
m_{0}^{2}L^{2}+j^{2}/r^{2}\right) \right]}.
\end{equation*}
And, $W_{\pm}(r)$ are computed as
\begin{equation}
W_{\pm }(r)=\pm i\pi \frac{E}{r_{+}-r_{-}}\left[ 1+\beta\Sigma\right],\label{MEqq2}
\end{equation}
where the abbreviation $\Sigma$ is
\begin{equation*}
\Sigma=\frac{\left( r_{+}-r_{-}\right) ^{2}\left[
3L^{2}M_{0}^{2}r_{+}^{2}+3j^{2}\right] +4E^{2}r_{+}^{2}}{2L^{2}r_{+}^{2}%
\left( r_{+}-r_{-}\right)^{4}},
\end{equation*}
where $W_{+}(r_{h})$ is outgoing and $W_{-}(r_{h})$ is incoming
solutions of the radial part. Then, using the Eq.(\ref{Equation16}), the tunneling probability of the scalar particle is calculated as
\begin{equation}
\Gamma =\exp \left(-\frac{4\pi E}{\hbar\left(
r_{+}-r_{-}\right) }\left[ 1+\beta \Sigma\right]
\right)\label{MEqq3}
\end{equation}
and, subsequently, using the Eq.(\ref{Equation18}), the modify Hawking temperature of the scalar particle, $T_{H}^{KG}$, becomes as
\begin{equation}
T_{H}^{KG}=\frac{\hbar\left( r_{+}-r_{-}\right) }{4\pi}\frac{1}{\left[ 1+\beta \Sigma\right]}.\label{MEqq4}
\end{equation}
Furthermore, as neglecting the higher order of the $\beta$ terms in the
expanding form of $T_{H}^{KG}$ in terms of the $\beta$, we find the modify
Hawking temperature of the black hole as follows;
\begin{equation}
T_{H}^{KG}\simeq\frac{\hbar\left( r_{+}-r_{-}\right) }{4\pi}\left[1-\beta \Sigma\right] \label{MEqq5}
\end{equation}
From these results, it can seen that the modified Hawking temperature is
related not only to the mass parameter of the black hole, but also to the AdS$_{3}$ radius, $L$, (and, hence, to the graviton mass) and angular momentum, energy and mass of
the tunneling scalar particle.
Furthermore, as can be seen from Eq.(\ref{MEqq13} and Eq.(\ref{MEqq5}), the
Hawking temperature probed by a Dirac
particle is higher than that of a scalar particle: $T_{H}^{D}$=$%
T_{H}^{KG}+\beta \frac{\hbar j^{2}m^{2}}{\pi r_{+}^{2}(r_{+}-r_{-})}$ for $%
M_{0}$=$m_{0}$ and $L^{2}$=$\frac{1}{2m^{2}}$. On the other hand, in the
case of $\beta =0$, the modify Hawking temperature reduced to the standard
temperature obtained by quantum tunneling process of the point particles
with spin-0, spin-1/2 and spin-1, respectively \cite{GY1,GY3}.

\section{Concluding remarks}\label{conc}

In this study, we investigated the quantum gravity effect on the tunneled
both spin-0 scalar and spin-1/2 Dirac particles from New-type black hole in
the context of 2+1 dimensional New Massive Gravity. For this, at first,
using the GUP relations, we modified the Klein-Gordon and Dirac equations that
describe the spin-0 scalar and spin-1/2 Dirac particles, respectively.
Then, using the Hamilton-Jacobi method, the tunneling probabilities of the
these particles are derived, and subsequently, the corrected Hawking
temperature of the black hole is calculated. We find that the
modified Hawking temperature not only depends on the black hole's properties,
but also depends on the emitted particle's mass, energy and total angular momentum.
Also, it is worth to mention that, the modified Hawking temperature depends on
mass of the graviton, i.e. quantum particle which mediates gravitational
radiation in the context of New Massive Gravity. As can be seen from Eq.(\ref{MEqq13}), the Hawking temperature of the Dirac particle increase by the
total angular momentum of the particle while it decreases by the energy and
mass of the particle and the graviton mass.

In addition, we can summarize some important results as follows:

\begin{itemize}
\item In Eq.(\ref{MEqq5}), the modify Hawking temperature of the scalar particle is lower than the standard Hawking temperature.

\item However, in Eq.(\ref{MEqq13}), as $4E^{2}r_{+}^{2}+\frac{3m_{0}^{2}}{2m^{2}}r_{+}^{2}(r_{+}-r_{-})^{2}<j^{2}(r_{+}-r_{-})^{2}$, the modify
Hawking temperature of the Dirac particle is higher than the standard
Hawking temperature. Furthermore, when $4E^{2}r_{+}^{2}+\frac{3m_{0}^{2}}{2m^{2}}r_{+}^{2}(r_{+}-r_{-})^{2}>j^{2}(r_{+}-r_{-})^{2}$, the modify
Hawking temperature is lower than the standard Hawking temperature. If $4E^{2}r_{+}^{2}+\frac{3m_{0}^{2}}{2m^{2}}r_{+}^{2}(r_{+}-r_{-})^{2}$=$j^{2}(r_{+}-r_{-})^{2}$, then the GUP effect is canceled, and the Hawking
temperature of the Dirac particle reduces to the standard Hawking temperature.

\item According to Eq.(\ref{MEqq13}) and Eq.(\ref{MEqq5}), the modify Hawking temperature of the New-type black hole probed by tunneling Dirac particle is higher than that of scalar particle:
\begin{eqnarray*}
T_{H}^{D}=T_{H}^{KG}+\beta \frac{\hbar j^2 m^2}{\pi r_{+}^2(r_{+}-r_{-})}.
\end{eqnarray*}
where we adopt that the mass of the Dirac particle is equivalent to the mass of the scalar particle, i.e. $m_{0}=M_{0}$ .
\item The New-type black hole is classified as six classes according to the signatures of the
parameters $b$ and $c$, and, hence, it exhibits different physical and mathematical properties. For example, it reduced to the static BTZ black hole in the case of $b=0$ and $c<0$.
In this context, according to tunneling of the scalar and Dirac particles, the modify Hawking temperature of the static BTZ black hole is
\begin{eqnarray*}
T_{H}^{KG}=T_{H}\left[ 1-\alpha m^{2} \frac{4 \left(\frac{3 M_{0}^{2}}{2m^{2}}\left\vert c\right\vert+3j^{2}\right) +4E^{2}}{c^{2}}\right]
\end{eqnarray*}
and
\begin{eqnarray*}
T_{H}^{D}=T_{H}\left[ 1-\alpha m^{2} \frac{4\left(\frac{3 m_{0}^{2}}{2m^{2}}\left\vert c\right\vert-j^{2}\right) +4E^{2}}{c^{2}}\right],
\end{eqnarray*}
respectively. Here, $r_{+}=-r_{-}=\sqrt{\left\vert c\right\vert}$ is used and the $T_{H}=\hbar \frac{\sqrt{\left\vert c\right\vert }}{2\pi }$ is the standard Hawking temperature of the static
BTZ black hole in the context of the 2+1 dimensional New Massive Gravity theory \cite{GY1}.
\item In the absence of the quantum gravity effect, i.e. $\beta$=$0$, the modify Hawking temperature is reduced
to the standard temperature obtained by quantum tunneling of the massive spin-0, spin-1/2 and spin-1 point particles \cite{GY1,GY3}.
\end{itemize}

Finally, in the context of GUP, we have seen that the graviton and the tunneling particle masses have an affect decreasing the Hawking temperature in both scalar and Dirac particle tunneling proses. On the other hand, the total angular momentum has different effect on the Hawking temperature for a type of tunneling particle. For a scalar particle, it results in decrease in the temperature whereas it provides an increase in the temperature for a Dirac particle. These results show that the intrinsic properties of the particle, except total angular momentum for the Dirac particle, and graviton mass may cause screening for the black hole radiation.

\section*{Acknowledgements}

This work was supported by Akdeniz University, Scientific Research Projects Unit.

\end{document}